\documentclass{llncs}

\usepackage{ajost}
\usepackage{graphicx}
\usepackage{amssymb}
\usepackage{color}
\usepackage{float}
\restylefloat{figure}

\usepackage[T1]{fontenc}

\usepackage{txfonts}
\usepackage{enumitem}
\usepackage{upgreek}

\definecolor{coolblack}{rgb}{0.0, 0.18, 0.39}

\newcommand{\COMMENT}[1]{}

\renewcommand{\tt}{\rm \ttfamily }
\newcommand{\codefont}{\tt}
\newcommand{\code}[1]{\mbox{\codefont{#1}}}
\newcommand{\us}{\raise-.8ex\hbox{-}}

\newcommand{\equcomment}[1]{\textnormal{\textit{#1}}}
\newcommand{\equprogram}[1]{%
\frenchspacing%
\refstepcounter{equation}%
\par\vspace\baselineskip\hspace{1em}%
\scalebox{0.93}{$\vcenter{\codefont\noindent{#1}}$}%
\raisebox{0.3ex}{\kern-0.4em\llap{\rm (\theequation)}}
\par\vspace\baselineskip\noindent\kern-.0em%
}

\newcommand{\equdef}[1]{%
\frenchspacing%
\par\vspace\baselineskip\hspace{1em}%
\scalebox{0.93}{$\vcenter{\codefont\noindent{#1}}$}%
\raisebox{0.3ex}{\kern-2.5em\llap{}}
\par\vspace\baselineskip\noindent\kern-.0em%
}

\def\boxit#1{\raisebox{-3.5pt}{\vbox{\hrule\hbox{\vrule\kern3pt
  \vbox{\kern3pt\hbox{#1}\kern3pt}\kern3pt\vrule}\hrule}}}

\usepackage{pstricks}

\newcommand{\sprite}{{\sc Sprite}}
\newcommand{\kics}{{\sc KiCS2}}
\newcommand{\pakcs}{{\sc Pakcs}}
\newcommand{\mcc}{{\sc Mcc}}

\begin{document}
\pagestyle{empty}  

\title{A New Functional-Logic Compiler for Curry: {\sprite}
\thanks{This material is based upon work partially supported by the National
Science Foundation under Grant No. CCF-1317249.}
}

\author{
  Sergio Antoy
  \and
  Andy Jost
}
\institute{
  Computer Science Dept., Portland State University, Oregon, U.S.A.
\\[1ex]
\email{antoy@cs.pdx.edu} \\
\email{ajost@pdx.edu}
}
\maketitle


\begin{abstract}
\sloppy
We introduce a new native code compiler for Curry codenamed {\sprite}. {\sprite} is
based on the Fair Scheme, a compilation strategy that provides instructions for
transforming declarative, non-deterministic programs of a certain class into
imperative, deterministic code.  We outline salient features of {\sprite}, discuss
its implementation of Curry programs, and present benchmarking results.  {\sprite}
is the first-to-date operationally complete implementation of Curry.
Preliminary results show that ensuring this property does not incur a
significant penalty.
\end{abstract}

\noindent
{\bf Keywords:}
Functional logic programming, Compiler implementation, Operational completeness


\section{Introduction}
\label{Introduction}
The functional-logic language Curry \cite{hanus06curry,Hanus13} is a
syntactically small extension of the popular functional language Haskell.  Its
seamless combination of functional and logic programming concepts gives rise to
hybrid features that encourage expressive, abstract, and declarative programs
\cite{AntoyHanus10CACM,Hanus13}.

One example of such a feature is a functional pattern \cite{antoyhanus05lopstr}, in which
functions are invoked in the left-hand sides of rules.  This is an
intuitive way to construct patterns with syntactically-sugared high-level
features that puts patterns on a more even footing with expressions.  In
Curry, patterns can be composed and refactored like other code, and
encapsulation can be used to hide details.  We illustrate this with function
\code{get}, defined below, which finds the values associated with a key in a
list of key-value pairs.
\equprogram{%
with x = \_ ++ [x] ++ \_ \\
get key (with (key, value)) = value
}
Operation \code{with} generates all lists containing \code{x}.
The anonymous variables, indicated by
``\code{\_\kern1pt}'', are place holders for expressions that are not used.
Function ``\code{++}'' is the list-appending operator.
When used in a left-hand side, as in the rule for \code{get},
operation \code{with}
produces a pattern that matches any list containing \code{x}.  Thus, the second
argument to \code{get} is a list --- any list --- containing the pair
\code{(key, value)}.  The repeated variable, \code{key}, implies a constraint
that, in this case, ensures that only values associated with the given key are
selected.

By similar means, we may identify keys:
\equprogram{%
key\_of (with (key, \_\kern1pt)) = key
}
This non-deterministically returns a key of the given list; for example:
\equprogram{%
> key\_of [('a',0), ('b',1), ('c',2)]\\
\noindent 'a'\\
\noindent 'b'\\
\noindent 'c'
}
This is just one of many features \cite{AntoyHanus10CACM,Hanus13}
that make Curry an appealing choice, particularly when the
desired properties of a program result
are easy to describe, but a set of step-by-step
instructions to obtain the result is more difficult to come by.

This paper describes work towards a new Curry compiler we call {\sprite}.  {\sprite}
aims to be the first operationally complete Curry compiler, meaning it should
produce all values of a source program (within time and space constraints).
Our compiler is based on a compilation strategy named the Fair Scheme
\cite{antoyjost13lopstr} that sets out rules for compiling a functional-logic
program (in the form of a graph rewriting system) into abstract
\emph{deterministic} procedures that easily map to the instructions of a
low-level programming language.  Section~\ref{Sprite} introduces {\sprite} at a
high level, and describes the transformations it performs.
Section~\ref{Implementation} describes the implementation of Curry programs as
imperative code.  Section~\ref{Performance} contains benchmark results.
Section~\ref{Related Work} describes other Curry compilers.
Section~\ref{Future Work} addresses future work, and
Section~\ref{Conclusion} contains our concluding remarks.


\section{The {\sprite} Curry Compiler}
\label{Sprite}

{\sprite} is a native code compiler for Curry.  Like all compilers, {\sprite}
subjects source programs to a series of transformations.  To begin, an external
program is used to convert Curry source code into a desugared representation
called FlatCurry \cite{flatcurry}, which {\sprite} further transforms into a
custom intermediate representation we call ICurry.  Then, following the steps
laid out in the Fair Scheme, {\sprite} converts ICurry into a graph rewriting
system that implements the program.  This system is realized in a low-level,
machine-independent language provided by the open-source compiler
infrastructure library LLVM \cite{LLVM:CGO04}.  That code is then optimized and
lowered to native assembly, ultimately producing an executable program.  {\sprite}
provides a convenience program, \code{scc}, to coordinate the whole procedure.

\subsection{ICurry}

ICurry, where the ``I'' stands for ``imperative,'' is a form of Curry programs
suitable for translation into imperative code.  ICurry is inspired by FlatCurry
\cite{flatcurry}, a popular representation of Curry programs that has been very
successful for a variety of tasks including implementations in Prolog
\cite{hanus14pakcs}.  FlatCurry provides expressions that resemble those of a
functional program --- e.g., they may include local declarations in the form of
let blocks and conditionals in the form of case constructs, all possibly
nested.  Although the pattern-matching strategy is made explicit through case
expressions, FlatCurry is declarative.  ICurry's purpose is to represent the
program in a more convenient imperative form --- more convenient since
{\sprite} will ultimately implement it in an imperative language.  In
imperative languages, local declarations and conditionals take the form of
statements while expressions are limited to constants and/or calls to
subroutines, possibly nested.  ICurry provides statements for local
declarations and conditionals.  It provides expressions that avoid constructs
that cannot be directly translated into the expressions of an imperative
language.

In ICurry all non-determinism --- including the implicit non-determinism in
high-level features, such as functional patterns --- is expressed through
choices.  A choice is the archetypal non-deter\-ministic function, indicated by
the symbol ``\code{?}'' and defined by the following rules:
\equprogram{%
x ? \_ = x\\
\_ ? y = y
}
The use of only choices is made possible, in part, by a duality between choices
and free variables \cite{AntoyHanus06ICLP,DiosCastro-LopezFraguas07ENTCS}: any
language feature expressed with choices can be implemented with free variables
and vice versa.  Algorithms exists to convert one to the other, meaning we are
free to choose the most convenient representation in {\sprite}.

Finally, as in FlatCurry, the pattern-matching strategy in ICurry is made
explicit and guided by a definitional tree \cite{antoy92alp}, a structure made
up of stepwise case distinctions that combines all rules of a function.  We
illustrate this for the \code{zip} function, defined as:
\equprogram{%
zip [] \_ = []\\
zip (\kern1pt\_\kern1pt:\kern1pt\_\kern1pt) [] = []\\
zip (x:xs) (y:ys) = (x,y) : zip xs ys
}
The corresponding definitional tree is shown below as it might appear in
ICurry.
\equprogram{\label{zip-dt}%
zip = \textbackslash a b -> case a of\\
\phantom{1234}[]\phantom{12345}-> []\\
\phantom{1234}(x:xs) -> case b of\\
\phantom{12345678}[]\phantom{12345}-> []\\
\phantom{12345678}(y:ys) -> (x,y) : zip xs ys
}

\subsection{Evaluating ICurry}
\label{evaluating}

It is understood how to evaluate the right-hand side of (\ref{zip-dt})
efficiently; the Spineless Tagless G-machine (STG) \cite{peyton1989spineless},
for instance, is up to the task.  But the non-deterministic properties of
functional-logic programs complicate matters.  To evaluate \code{zip}, its
first argument must be reduced to head-normal form.  In a purely functional
language, the root node of a head-normal form is always a data constructor
symbol (assuming partial application is implemented by a data-like object),
or else the computation fails.  But for functional-logic programs, two
additional possibilities must be considered, leading to an extended case
distinction:
\equprogram{%
\label{zip-ext}
zip = \textbackslash a b -> case a of\\
\phantom{1234}x ? y\phantom{12}-> \equcomment{(pull-tab)}\kern19.5pt\equcomment{- - implied}\\
\phantom{1234}$\bot$\kern30pt -> $\bot$\kern50pt\equcomment{- - implied}\\
\phantom{1234}[]\phantom{12345}-> []\\
\phantom{1234}(x:xs) -> case b of ...
}
The infrastructure for executing this kind of pattern matching very efficiently
by means of dispatch tables will be described shortly, but for now we note two
things.  First, there is no need for ICurry to spell out these extra cases, as
they can be generated by the compiler.  Second, their presence calls for an
expanded notion of the computation that allows for additional node states.
Because of this, {\sprite} hosts computations in a graph whose nodes are taken
from four classes: \emph{constructors}, \emph{functions}, \emph{choices}, and
\emph{failures}.  Constructors and functions are provided by the source
program; choices are built-in; and failures, denoted ``$\bot$'', arise from
incompletely defined operations such as \code{head}, the function that returns
the head of a list.  For example, \code{head\,[]} rewrites to ``$\bot$''.  A
simple replacement therefore propagates failure from needed arguments to roots.

Choices execute a special step called a \emph{pull-tab}
\cite{antoy11iclp,BrasselHuchAPLAS07}.
Pull-tab steps lift non-det\-er\-minism out of needed positions, where they prevent
completion of pattern matches.  The result is a choice between two
more-definite expressions.  A pull-tab step is shown below:
\equprogram{%
\label{pull-tab}
zip (a ? b) c $\rightarrow$ zip a x ? zip b x where x = c
}
A pattern match cannot proceed while \code{(a ? b)} is the first argument to
\code{zip} because there is no matching rule in the function definition (one
cannot exist because the choice symbol is disallowed on left-hand sides).  We
do not want to choose between \code{a} and \code{b} because such a choice would
have to be reconsidered to avoid losing potential results.  The pull-tab
transformation ``pulls'' the choice to an outermore position, producing two
new subexpressions, \code{zip a c} and \code{zip b c}, that can be evaluated
further.  The fact that \code{c} is shared in the result illustrates a
desirable property: that node duplication is minimal and localized.
Pull-tabbing involves some technicalities that we address later.  The complete
details are in \cite{antoy11iclp}.

Due to the extra cases, additional node types, and, especially, the unusual
mechanics of pull-tabbing steps, we chose to develop in {\sprite} a new
evaluation machine from scratch rather than augment an existing one such as
STG.  The property of pull-tabbing that it ``breaks-out'' of
recursively-descending evaluation into nested expressions fundamentally changes
the computation so that existing functional strategies are difficult to apply.
In {\sprite}, we have implemented \emph{de novo} an evaluation mechanism and
runtime system based on the Fair Scheme.  These are the topic of the next
section.


\section{Implementation}
\label{Implementation}

In this section, we describe the implementation of Curry programs in imperative
code.  {\sprite} generates LLVM code, but we assume most readers are not familiar
with that.  So, rather than presenting the generated code, we describe the
implemented programs in terms of familiar concepts that appear directly in
LLVM.  In this way, the reader can think in terms of an unspecified target
language --- one similar to assembly --- that implements those concepts.  To
facilitate the following description, we indicate in parentheses where a
similar feature exists in the C programming language.

In the target language,
values are strongly typed, and the types include integers, pointers, arrays,
structures and functions.  Programs are arranged into compilation units called
modules that contain symbols.  Symbols are visible to other modules, and to
control access to them each one is marked internal (\code{static}) or external
(\code{extern}).  Control flow within functions is carried out by branch
instructions.  These include unconditional branches (\code{goto}), conditional
branches (\code{if}, \code{for}, \code{while}) and indirect branches
(\code{goto*}).  The target of every branch instruction is a function-local
address (label).  A call stack is provided, and it is manipulated by call and
return instructions that enter and exit functions, respectively.  Calls are
normally executed in a fresh stack frame, but the target language also supports
explicit tail recursion, and {\sprite} puts it to good use.

\subsection{Expression Representation}

The expressions evaluated by a program are graphs consisting of labeled
nodes having zero or more successors.  Each node belongs to one of four
classes, as discussed in the previous section.  For constructors and functions,
node labels are equivalent to symbols defined in the source program.  Failures
and choices are labeled with reserved symbols.  Successors are references to
other nodes.  The number of successors, which equals the arity of the
corresponding symbol, is fixed at compile time.  Partial applications are
written in \emph{eval/apply} form \cite{marlow2004making}.

\begin{figure}
\centering
\caption{The heap object layout.\label{heap-object}}
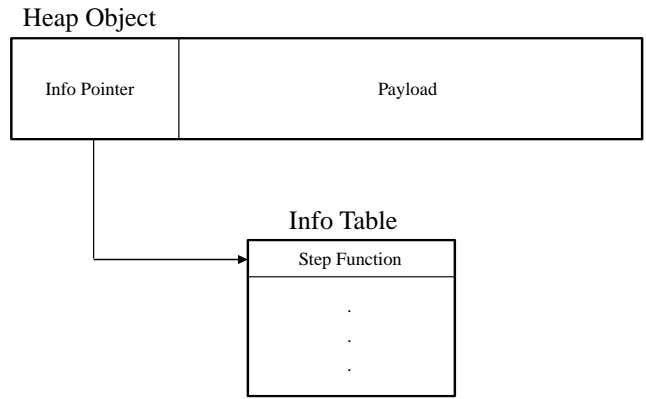
\end{figure}
{\sprite} implements graph nodes as heap objects.  The layout of a heap object is
shown in Fig.~\ref{heap-object}.  The label is implemented as a pointer to a
static info table that will be described later.  {\sprite} emits exactly one table
for each symbol in the Curry program.  Successors are implemented as pointers
to other heap objects.

\subsection{Evaluation}
\label{evaluation}

Evaluation in {\sprite} is the repeated execution of rewriting and pull-tabbing
steps.  Both are implemented by two interleaved activities:
replacement and pattern-matching.  A replacement produces a new graph from a
previous one by replacing a subexpression matching the left-hand side of a rule
with the corresponding right-hand side.  For instance, $1+1$ might be replaced
with $2$.  A replacement is implemented by overwriting the heap object at the
root of the subexpression being replaced.  The key advantage of this
destructive update is that no pointer redirection \cite[Def.
8]{EchahedJanodet97IMAG}\cite{Glauert97JPL} is required during a rewrite step.
Reusing a heap object also has the advantage of saving one memory allocation
and deallocation per replacement, but requires that every heap object be
capable of storing any node, whatever its arity.  {\sprite} meets this
requirement by providing in heap objects a fixed amount of space capable of
holding a small number of successors.  For nodes with more successors than
would fit in this space, the payload instead contains a pointer to a larger
array.  This approach simplifies memory management for heap objects: since they
are all the same size, a single memory pool suffices.  Because arities are
known at compile time, no runtime checks are needed to determine whether
successor pointers reside in the heap object.

Pattern-matching consists of cascading case distinctions over the root symbol
of the expression being matched that culminate either in a replacement or in
the patter match of a subexpression.  The Fair Scheme implements this according
to a strategy guided by the definitional trees encoded in ICurry.  Case
distinction as exemplified in~(\ref{zip-ext}) assumes that an expression being
matched is not rooted by a function symbol.  Thus, when a node needed to
complete a match is labeled by a function symbol, the expression rooted by that
node is evaluated until it is labeled by a non-function symbol.  A
function-labeled node, $n$, is evaluated by a target function called the \emph{step}
function that performs a pattern match and replacement at $n$.  Each Curry
function gives rise to one target function, a pointer to which is stored in the
associated info table (see Fig.~\ref{heap-object}).
\newcommand{\tabentry}[3]{%
  \rput[l](2,#3){#1 \ \ \psframebox[linecolor=black]{\phantom{X;}}}
  \psline{->}(2.7,#3)(3.4,#3)
  \rput[l](3.6,#3){#2}
}
\begin{figure}[!hb]
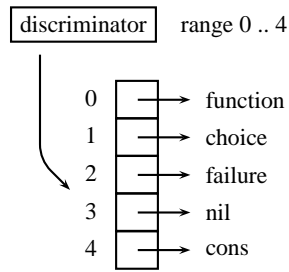

  \hrulefill\\
  \centering
  \pspicture(0,3.4)(6,7.8)
  \rput[l](1,7){\psframebox[linecolor=black]{discriminator} \ \ \ range 0 .. 4}
  \psline[linearc=0.5]{->}(1.4,6.6)(1.4,5.2)(1.8,4.8)

  \tabentry{0}{function}{6}
  \tabentry{1}{choice}{5.5}
  \tabentry{2}{failure}{5}
  \tabentry{3}{nil}{4.5}
  \tabentry{4}{cons}{4}

  \endpspicture
  \\\hrulefill\\
\caption{
  \label{jump-table-scheme}{Schematic representation of the {\sprite}
  tagged dispatching mechanism for a distinction of a \code{List} type.}
}
\end{figure}

Operationally, pattern-matching amounts to evaluating nested case expressions
similar to the one shown in~(\ref{zip-ext}).  {\sprite} implements this through
a mechanism we call \emph{tagged dispatch}.  With this approach, the compiler
assigns each symbol a tag at compile time.  Tags are sequential integers
indicating which of the four classes discussed earlier the node belongs to.
The three lowest tags are reserved for functions, choices, and failures (all
functions have the same tag).  For constructors, the tag additionally indicates
\emph{which} constructor of its type the symbol represents.  To see how this
works, consider the following type definition:
\equprogram{%
data ABC = A | B | C
}
\code{ABC} comprises three constructors in a well-defined order (any fixed
order would do).  To distinguish between them, {\sprite} tags these with
sequential numbers starting at the integer that follows the reserved tags.
So, the tag of \code{A} is one less than the tag of \code{B}, which is one less
than the tag of \code{C}.  These values are unique within the type, but not
throughout the program: the first constructor of each type, for instance,
always has the same tag.  Following these rules, it is easy to see that every
case discriminator is a node tagged with one of $3+N$ consecutive integers,
where $N$ is the number of constructors in its type.  To compile a case
expression, {\sprite} emits a jump table that transfers control to a code block
appropriate for handling the discriminator tag.  For example, the block that handles
failure rewrites to failure, and the block that handles choices executes a
pull-tab.  This is shown schematically in Fig.~\ref{jump-table-scheme}.  It is
in general impossible to know at compile time which constructors may be
encountered when the program runs, so the jump table must be complete.  If a
functional logic program does not define a branch for some constructor ---
i.e., a function is not completely defined --- the branch for that constructor
is a rewrite to failure.
\begin{figure}[!b]
\centering
\hrulefill\\
\parbox{11cm}{
\equdef{%
static void* jump\_table[5] = \{\\
\phantom{1234}\&function\_tag, \&choice\_tag, \&failure\_tag, \&nil\_tag, \&cons\_tag\\
\phantom{12}\};\\
\\
entry:\phantom{1234567890}goto* jump\_table[discriminator.tag];\\
function\_tag: \phantom{12}call\_step\_function(discriminator);\\
\phantom{1234567890123456}goto* jump\_table[discriminator.tag];\\
choice\_tag:\phantom{12345}/*execute a pull tab*/\\
failure\_tag:\phantom{1234}/*rewrite to failure*/\\
nil\_tag:\phantom{12345678}/*rewrite to []*/\\
cons\_tag:\phantom{1234567}/*process the nested case expression*/\\
}}\\
\hrulefill\\
\caption{
\label{jump-table}
An illustrative implementation in C of the case expression shown
in~(\ref{zip-ext}).  This code fragment would appear in the body of the step
function for \code{zip}.  Variable \code{discriminator} refers to the
case discriminator.  Label \code{entry} indicates the entry point into this
case expression.}
\end{figure}

To implement tagged dispatch, {\sprite} creates function-local code blocks as
labels, constructs a static jump table containing their addresses, and executes
indirect branch instructions --- based on the discriminator tag --- through the
table.  Figure~\ref{jump-table} shows a fragment of C code that approximates
this.  Case distinction occurs over a variable of \code{List} type with two
constructors, \code{nil} and \code{cons}.  Five labeled code blocks handle the
five tags that may appear at the case discriminator.  A static array of label
address implements the jump table.  This example assumes the \emph{function},
\emph{choice}, \emph{failure}, \emph{nil}, and \emph{cons} tags take the values
zero through four, respectively.  The jump table contains one extra case not
depicted in (\ref{zip-ext}).  When the discriminator is a \emph{function}, the
step function of the discriminator root label is applied as many times as
necessary until the discriminator class is no longer \emph{function}.


\subsection{Completeness and Consistency}
\label{completeness}

{\sprite} aims to be the first complete Curry compiler.  Informally, complete
means the program produces all the intended results of the source program.
More precisely, and especially for infinite computations, an arbitrary value
will eventually be produced, given enough resources.  This is a difficult
problem because a non-terminating computation for obtaining one result could
block progress of some other computation that would obtain another result.  For
example, the following program has a result, 1, that can be obtained in only a
couple of steps, but existing Curry compilers fail to produce it:
\equprogram{\label{loop}%
loop = loop\\
main = loop ? (1 ? loop)
}
The Fair Scheme defines a complete evaluation strategy.  It creates a
work queue containing all of the expressions that might produce a result.
At all
times, the expression at the head of the queue is active, meaning it is
being evaluated.  Initially, the work queue contains only the goal expression.
Whenever pull-tabbing places a choice at the root of an expression, that
expression forks.  It is removed from the queue, and its two alternatives are
added.  Whenever an expression produces a value, it is removed from the queue.
To avoid endlessly working on an infinite computation, the program rotates the
active computation to the end of the work queue every so often.  In so doing,
{\sprite} guarantees that no expression is ignored forever, hence
no potential result is lost.

A proof of correctness of compiled programs
is provided in \cite{antoyjost13lopstr}
for the abstract formulation of the compiler, the Fair Scheme.
In this domain, correctness is the property that an executable
program produces all and only the values intended
by the corresponding source program.
A delicate point is raised by pull-tabbing.
A pull-tab step may duplicate or clone a choice, as the following example shows.
Cloned choices should be seen as a single choice.
Thus when a computation reduces a choice to its right alternative,
it should also reduce any other clone of the same choice
to the right alternative, and likewise for the left alternative.
Computations obeying this condition are called \emph{consistent}.
\equprogram{%
xor x x where x = T ? F\\
\phantom{1234}$\rightarrow_{pull-tab}$ (xor T x) ? (xor F x) where x = T ? F\\
}
In the example above, a pull-tab step applied to the choice in \code{x} leads
to its duplication.  Now, when evaluating either alternative of the topmost
choice, a consistent strategy must recognize that the remaining choice (in
\code{x}) is already made.  For instance, when evaluating \code{xor T x}, the
value of \code{x} can only be \code{T}, the left alternative, because
the left alternative of \code{x} has already been selected to obtain
\code{xor T x}.
To keep track of clones, the Fair Scheme annotates choices with
identifiers.  Two choice nodes with identical identifiers represent the same
choice.  Fresh identifiers are assigned when new choices arise from a
replacement; pull-tab steps copy existing identifiers.  Every expression in
the work queue owns a fingerprint, which is a mapping from choice identifiers
to values in the set \{\emph{left,right,either}\}.
The fingerprint is used to
detect and remove inconsistent computations from the work queue.

It is possible to syntactically pre-compute pull-tab steps: that is, a case
statement such as the one in (\ref{zip-ext}) could implement pull-tabbing by
defining an appropriate right-hand side rule for the choice branch.  In fact, a
major competing implementation of Curry does exactly that
\cite{brasselhanuspeemoellerreck11}.  A disadvantage of that approach is that
choice identifers must appear as first-class citizens of the program and be
propagated through pull-tab steps using additional rules not encoded in the
source program.  We believe it is more efficient to embed choice identifiers in
choice nodes as an implementation detail and process pull-tab steps
dynamically.  Section~\ref{flp-perf} compares these two approaches in greater
detail.


\section{Performance}
\label{Performance}

In this section we present a set of benchmark results.  These programs were
previous used to compare three implementations of Curry
\cite{brasselhanuspeemoellerreck11}: \mcc, \pakcs, and \kics.  We shall use
\kics\ to perform direct comparisons with {\sprite}\footnote{Available at
\url{https://github.com/andyjost/Sprite-3} }, since it compares favorably to
the others, and mention the relative performance of the others.  \kics\
compiles Curry to Haskell and then uses the Glasgow Haskell Compiler (GHC)
\cite{ghchomepage} to produce executables.  GHC has been shown to produce very
efficient code \cite{jones1995compilation,jones1996compiling,partain1993nofib}.
Like {\sprite}, \kics\ uses a pull-tabbing evaluation strategy, but unlike
{\sprite}, it does not form a work queue; hence, is incomplete when faced with
programs such as (\ref{loop}).  Instead, it builds a tree containing all values
of the program and executes (lazily and with interleaved steps) a user-selected
search algorithm.

A major highlight of \kics\ is that purely functional programs compile to
``straight'' Haskell, thus incurring no overhead due to the presence of unused
logic capabilities.  {\sprite}, too, enjoys this zero-overhead property, but
there is little room to improve upon GHC for functional programs, as it is the
beneficiary of exponentially more effort.  Our goal for functional programs,
therefore, is simply to measure and minimize the penalty of running {\sprite}.
For programs that utilize logic features \kics\ emits Haskell code that
simulates non-determinism. In these cases, there is more room for improvement
since, for example, {\sprite} can avoid simulation overhead by more directly
implementing logic features.

\setlength{\tabcolsep}{8pt}
\begin{figure}[!t]
\centering
\begin{tabular}{llrrr}
  \hline
  Program & Type & \kics & {\sprite} & $\Updelta$\kern0.3em\\
  \hline
  PaliFunPats & FL  & 0.64 & 0.09 & -7.1\\
  LastFunPats & FL  & 1.85 & 0.30 & -6.2\\
  Last  & FL  & 1.90 & 0.31 & -6.1\\
  PermSortPeano & FL  & 44.04 & 8.14 & -5.4\\
  PermSort  & FL  & 42.72 & 8.15 & -5.3\\
  ExpVarFunPats & FL  & 5.92 & 1.29 & -4.6\\
  Half  & FL  & 42.31 & 9.55 & -4.4\\
  Reverse & F & 0.36 & 0.21 & -1.7\\
  ReverseUser & F & 0.34 & 0.21 & -1.6\\
  ReverseBuiltin  & F & 0.40 & 0.39 & -1.0\\
  ReverseHO & F & 0.36 & 0.39 & 1.1\\
  Primes  & F & 0.29 & 0.32 & 1.1\\
  ShareNonDet & FL  & 0.28 & 0.33 & 1.2\\
  PrimesBuiltin & F & 0.73 & 1.10 & 1.5\\
  PrimesPeano & F & 0.41 & 0.66 & 1.6\\
  QueensUser  & F & 0.87 & 1.83 & 2.1\\
  Queens  & F & 0.80 & 1.81 & 2.3\\
  TakPeano  & F & 0.84 & 2.08 & 2.5\\
  Tak & F & 0.32 & 0.92 & 2.9\\
  \hline
\end{tabular}
\caption{\label{benchmarks}Execution times for a set of functional ($F$) and
functional-logic ($FL$) programs taken from the \kics\ benchmark suite.  Times
are in seconds.  The final column ($\Updelta$) reports the speed-up (negative) or
slow-down (positive) factor of {\sprite} relative to \kics.  System configuration:
Intel i5-3470 CPU at 3.20GHz, Ubuntu Linux 14.04.}
\end{figure}

\subsection{Functional Programs}
The execution times for a set of programs taken from the \kics\ benchmark
suite\footnote{Downloaded from
\url{https://www-ps.informatik.uni-kiel.de/kics2/benchmarks.}} are shown in
Fig.~\ref{benchmarks}.  The results are arranged in order from greatest
improvement to greatest degradation in execution time.  The most striking
feature is the clear division between the functional (deterministic) and
functional-logic (non-deterministic) subsets, which is consistent with our
above-stated expectations.
On average, {\sprite} produces relatively slower code
for functional programs and relatively faster code for functional-logic ones.
We calculate averages as the geometric mean, since that method is not
strongly influenced by extreme results in either direction.  The functional
subset runs, on average, 1.4\texttt{x}\ slower in {\sprite} as compared to
\kics.  Figures published by Bra{\ss}el et al.~\cite[Fig.2,
Fig.3]{brasselhanuspeemoellerreck11} indicate that \pakcs\ and \mcc\ run
148\texttt{x}\ and 9\texttt{x}\ slower than \kics, respectively, for these
programs.  We take these results as an indication that the functional parts
of {\sprite} --- i.e., those parts responsible for pattern-matching, rewriting,
memory management, and optimization --- although not as finely-tuned as their
GHC counterparts, still compare favorably to most mainstream Curry compilers.

We note that {\sprite} currently does not perform optimizations such as
deforestation \cite{gill1993short} or unboxing \cite{jones1995compilation}.
These, and other optimizations of ICurry, e.g.,
\cite{AntoyJohannsenLibby15EPTCS}, could potentially impact the benchmark
results.  Inspecting the output of GHC reveals that the \texttt{tak} program
(incidentally, the worse-case for {\sprite}) is optimized by GHC to a
fully-unboxed computation.  To see how LLVM stacks up, we rewrote the program
in C and converted it to LLVM using Clang \cite{clanghomepage}, a C language
front-end for LLVM.  When we compiled this to native code and measured the
execution time, we found that it was identical\footnote{Using the Linux
\texttt{time} command, whose resolution is 0.01 seconds.} to the \kics\ (and
GHC) time.  We therefore see no fundamental barrier to reducing the {\sprite}
``penalty'' to zero for this program, and perhaps others, too.  We have reason
to be optimistic that implementing more optimizations at the source and ICurry
levels, without fundamentally changing the core of {\sprite}, will yield
substantive improvements to {\sprite}.

\subsection{Functional-Logic Programs}
\label{flp-perf}
\begin{figure}[!ht]
\centering
\caption{\label{permsort-complexity}%
  Complexity analysis of \texttt{PermSort}.  Execution times are shown for a
  range of problem sizes.  The horizontal axis indicates the number of integers
  to sort by the permute-and-test method. 
}
\includegraphics[scale=0.8]{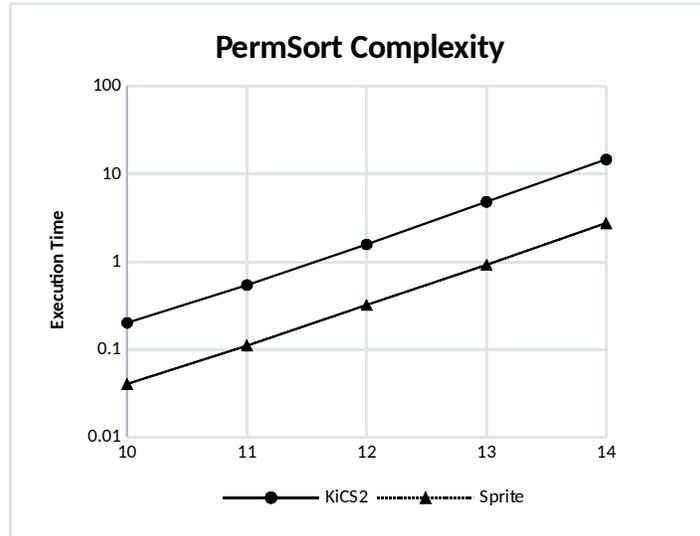}
\end{figure}

For the functional-logic subset, Fig.~\ref{benchmarks} shows that {\sprite}
produces relatively faster code: 4.4\texttt{x} faster, on average.  Published
comparisons \cite[Fig.4]{brasselhanuspeemoellerreck11} indicate that, compared
to \kics, \pakcs\ is 5.5\texttt{x}\ slower and \mcc\ is 3.5\texttt{x}\
\emph{faster} for these programs.  Our first thought after seeing this result
was that {\sprite} might enjoy a better algorithmic complexity.  We had just
completed work to reduce {\sprite}'s complexity when processing choices, so
perhaps, we thought, in doing that work we had surpassed \kics.  We set out to
test this by selecting a program dominated by choice generation and running it
for different input sizes, with and without the recent modifications to
{\sprite}.  The results are shown in Fig.~\ref{permsort-complexity}.
Contrary to our expectation, {\sprite} and \kics\ exhibit strikingly similar
complexity: both fit an exponential curve with $r^2$ in excess of 0.999, and their
slope coefficients differ by less than 2\%.  A better explanation, then, for the
difference is that some constant factor $c$ exists, such that choice-involved
steps in {\sprite} are $c$-times faster than in \kics.  What could account for
this factor?  We believe the best explanation is the overhead of simulating
non-determinism in Haskell, which we alluded to at the end of Sect.~\ref{completeness}.
To see why, we need to look at \kics\ in more detail.

\kics\ uses a few helper functions
\cite[Sect.~3.1]{brasselhanuspeemoellerreck11} to generate choice
identifiers:
\equprogram{%
thisID\phantom{aaaaaa}:: IDSupply -> ID\\
leftSupply\phantom{aa}:: IDSupply -> IDSupply\\
rightSupply\phantom{a}:: IDSupply -> IDSupply
}
The purpose of these functions is to ensure that choice identifiers are never
reused.  Here, \texttt{ID} is the type of a choice identifier and
\texttt{IDSupply} is opaque (for our purposes).  Any function that might
produce a choice is implicitly extended by \kics\ to accept a supply function.
As an example, this program
\equprogram{%
f :: Bool\\
main = xor f (False ? True)
}
is compiled to
\equprogram{%
main s = let s1 = leftSupply s\\
\phantom{aaaaaaaaaaaaa}s2 = rightSupply s\\
\phantom{aaaaaaaaaaaaa}s3 = leftSupply s2\\
\phantom{aaaaaaaaaaaaa}s4 = rightSupply s2\\
\phantom{aaaaa}in xor (f s3) (Choice (thisID s4) False True) s1
}
Clearly, the conversion to Haskell introduces overhead.  The point here is
simply to see that the compiled code involves five calls (to helper functions)
that were not present in the source program.  These reflect the cost of
simulating non-determinism in a purely-functional language.

In {\sprite}, fresh choice identifiers are created by reading and incrementing
a static integer.  Compared to the above approach, fewer parameters are passed
and fewer functions are called.  A similar approach could be used in a Haskell
implementation of Curry, but it would rely on impure features, adding another
layer of complexity and perhaps interfering with optimizations.  By contrast,
the \sprite\ approach is extreme in its simplicity, as it executes only a few
machine instructions.
There is a remote possibility that a computation could exhaust the supply of
identifiers since the type integer is finite.  \kics\ uses a list structure for
choice identifiers and so does not suffer from this potential shortcoming.
Certainly, the choice identifiers could be made arbitrarily large, but doing so
increases memory usage and overhead.  A better approach, we believe, would be
to compact the set of identifiers during garbage collection.  The idea is that
whenever a full collection occurs, {\sprite} would renumber the $n$ choice
identifiers in service at that time so that they fall into the contiguous range
$0,\ldots,n-1$.
This potential optimization illustrates the benefits of having total control
over the implementation, since in this case it makes modifying the garbage
collector a viable option.


\section{Related Work}
\label{Related Work}

Several Curry compilers are easily accessible, most notably \pakcs\
\cite{hanus14pakcs}, \kics\ \cite{brasselhanuspeemoellerreck11}
and \mcc\ \cite{luxmcc12}.
All these compilers implement a lazy evaluation strategy, based on
definitional trees, that executes only needed steps, but
differ in the control strategy that selects the order
in which the alternatives of a choice are executed.

Both \pakcs\ and \mcc\ use backtracking.
They attempt to evaluate all the values of the left
alternative of a choice before turning to the right alternative.
Backtracking is simple and relatively efficient,
but incomplete.  Hence, a benchmark against these compilers
may be interesting to understand the differences between
backtracking and pull-tabbing, but not to assess the
efficiency of {\sprite}.

By contrast, \kics's control strategy uses pull-tabbing, hence the computations
executed by \kics\ are much closer to those of {\sprite}.  \kics's compiler
translates Curry source code into Haskell source code which is then processed
by GHC \cite{ghchomepage}, a mainstream Haskell compiler.  The compiled code
benefits from a variety of optimizations available in GHC.
Section~\ref{Performance} contains a more detailed comparison between {\sprite} and
\kics.

There exist other functional logic languages,
e.g., ${\cal TOY}$ \cite{ToyHomepage,Lopez-FraguasSanchez-Hernandez99RTA},
whose operational semantics can be abstracted by needed narrowing
steps of a constructor-based graph rewriting system.
Some of our ideas could be applied with almost no changes
to the implementation of these languages.

A comparison with graph machines for functional languages is
problematic at best. 
Despite the remarkable syntactic similarities,
Curry's syntax extends Haskell's with a single construct,
a free variable declaration, the semantic differences are profound.
There are purely functional programs whose execution
produces a result as Curry, but does not terminate as
Haskell \cite[Sect. 3]{AntoyHanus10CACM}.
Furthermore, functional logic computations must be prepared
to encounter non-determinism and free variables.
Hence, situations and goals significantly differ.


\section{Future Work}
\label{Future Work}

Compilers are among the most complex software artifacts.
They are often bundled with extensions and additions such as
optimizers, profilers, tracers, debuggers, external libraries
for application domains such as databases or graphical user interfaces.
Given this reality, there are countless opportunities for future work.
We have no plans at this time to choose any one of the extensions and
additions listed above before any other.  Some optimizations mentioned
earlier, e.g., unboxing integers, are appealing only because they
would improve some benchmark, and thus the overall perceived performance of the
compiler, but they may contribute very marginally to the efficiency of
more realistic programs.
Usability-related extensions and additions, such as aids for tracing
and debugging an execution, and external libraries may better
contribute to the acceptance of our work.


\section{Conclusion}
\label{Conclusion}

We have presented {\sprite}, a new native code compiler for Curry.
{\sprite} combines the best features of existing Curry compilers.
Similar to {\kics}, {\sprite}'s strategy is based on pull-tabbing,
hence there is no an inherent loss of completeness of compilers based on
backtracking such as {\pakcs} and {\mcc}.
Similar to {\mcc}, {\sprite} compiles to an imperative target
language, hence is amenable to low-level machine optimization.
Differently from all existing compilers, {\sprite} is designed
to ensure operational completeness---all the values of an
expression are eventually produced given enough computational
resources.

{\sprite}'s main intermediate
language, ICurry, represents programs as graph rewriting systems.  We described
the implementation of Curry programs in imperative code using concepts of a
low-level target language.  Graph nodes are represented in memory as heap
objects, and an efficient mechanism called tagged dispatch is used to perform
pattern matches.  Finally, we discussed the mechanisms used by
{\sprite} to ensure completeness and consistency, and presented empirical data for
a set of benchmarking programs.  The benchmarks reveal that {\sprite} is
competitive with a leading implementation of Curry.


\bibliographystyle{plain}
\bibliography{biblio}

\end{document}